\documentclass[showpacs,aps,eqsecnum]{revtex4}
\usepackage{psfig,epsfig,graphicx,float,pst-all, rotating}

\begin{document}

\title{Phase transitions in the Interacting Boson Fermion Model: the gamma-unstable case}

\author{C.E. Alonso$^1$, J.M. Arias $^1$, L. Fortunato$^2$, and A. Vitturi$^2$ \\}

\affiliation{$^1$ Departamento de F\'{\i}sica At\'omica, Molecular y
  Nuclear, Facultad de F\'{\i}sica \\ Universidad de Sevilla, Apartado~1065, 
41080 Sevilla, Spain \\
$^2$ Dipartimento di Fisica Galileo Galilei and INFN,
Via Marzolo 8, 35131 Padova, Italy}


\begin{abstract}
The phase transition around the critical point in the evolution from
spherical to deformed $\gamma$-unstable shapes
is investigated in odd nuclei within the Interacting Boson
Fermion Model. 
We consider the particular case of an odd j=3/2 particle coupled to an even-even boson
core that undergoes a transition from spherical U(5) to
$\gamma$-unstable O(6) situation. The particular choice of the j=3/2 orbital
preserves in the odd case the condition of gamma-instability of the system.
As a consequence, energy spectrum and electromagnetic transitions, in correspondence 
of the critical point, display behaviours qualitatively similar to those of the
even core. The results are also in qualitative agreement with the recently proposed
E(5/4) model, although few differences are present, due to the different nature
of the two schemes.
\pacs{21.60.Re, 21.60.Fw}  
\end{abstract}

\maketitle

The study of phase transitions has recently received particular attention
in nuclear structure. The concept of critical point symmetry has been
first proposed in a number of cases by Iachello\cite{E5,X5,Y5}. These symmetries apply 
when a quantal system undergoes transitions between traditional
dynamical symmetries, as for example those characterizing situations
described in terms of harmonic vibrations 
or rigid rotations. Although these symmetries have been obtained
within the formalism based on the Bohr Hamiltonian\cite{BMII},  
their concept has also been used in connection with the Interacting
Boson Model (IBM)\cite{IBM}.

One of these critical point symmetries is associated with
the transition between spherical and $\gamma$-unstable shapes. Within
the IBM this can be obtained, for example, from the
Hamiltonian
\begin{equation}
H_B= x \hat n_d - \frac{1-x}{N} \hat Q_B \cdot \hat Q_B ~,
\label{HPP}
\end{equation}
which produces, varying the parameter $x$ from 1 to 0, a transition between the
two extreme situations characteristic of U(5)
and O(6) symmetries. The corresponding second-order shape phase transition has been 
investigated within the Bohr collective model in Ref. \cite{TR05}.
The operators appearing in the Hamiltonian above are given by
\begin{eqnarray}
\hat n_d & = & \sum_\mu d^\dag _\mu d_\mu ~~, \\
\hat Q_B  & = & (s^\dag \times \tilde d + d^\dag \times \tilde s)^{(2)} ~~,
\label{QB}
\end{eqnarray}
and $N$ is the total number of bosons.
For any value of $x$ this Hamiltonian maintains the typical degeneracies of the 
O(5) symmetry. Consistently with this,     
within the IBM coherent state formalism \cite{GK80,DSI80,BM80}, this Hamiltonian always
produces an energy surface which is independent of the $\gamma$ degree of freedom.  
In the $\beta$ 
variable, the energy surface displays a spherical minimum in $\beta=0$ for $x$ larger
than the critical value $x_c=\frac{4N-8}{5N-8}$, while having a deformed minimum
for values of $x$ smaller than the critical value.  At the critical point, the energy
surface acquires a $\beta^4$ behaviour \cite{beta4,beta4bis}, 
which is approximated by an infinite square well in the 
E(5) critical point symmetry \cite{E5} within the framework of the
collective Bohr Hamiltonian. 

Recently Iachello has discussed a supersymmetrical extension
of this concept, introducing the so-called E(5/4) model, where the boson part has a 
$\gamma-$independent square well potential and the boson-fermion coupling is
taken as a Spin(5) scalar interaction \cite{I05}. This solution describes
the spectral properties of odd-even nuclei at the transition between spherical and 
$\gamma-$unstable shapes. In this letter we want to discuss the evolution of odd nuclei and 
the corresponding behaviour at the critical point,
within the framework of the Interacting
Boson Fermion Model (IBFM) \cite{IVI91}, in which a single fermion is coupled to the even-even
bosonic core. The system will then be described by the Hamiltonian
\begin{equation}
H~=~H_{B}+H_{F}+V_{BF}~,
\label{HBF}
\end{equation}
where the term $V_{BF}$ couples the bosonic and fermionic parts.  In our case we
will assume the boson Hamiltonian to be of the form given in Eq. (\ref{HPP}).  
For the fermion
and boson-fermion parts we will take the particular choice of a particle moving in a 
single j-shell $j=3/2$ and a coupling term of the form
\begin{equation}
V_{BF}=- 2 \frac{1-x}{N} \hat Q_B \cdot \hat q_F~, 
\label{VBF}
\end{equation}
where $\hat Q_B$ (taken of the form given in Eq.(\ref{QB})) and
$\hat q_F= (a^\dag _{3/2} \times \tilde a_{3/2})^{(2)} $ are the  
boson and fermion quadrupole operators, respectively. 

\begin{figure}[!t]
\begin{center}
\epsfig{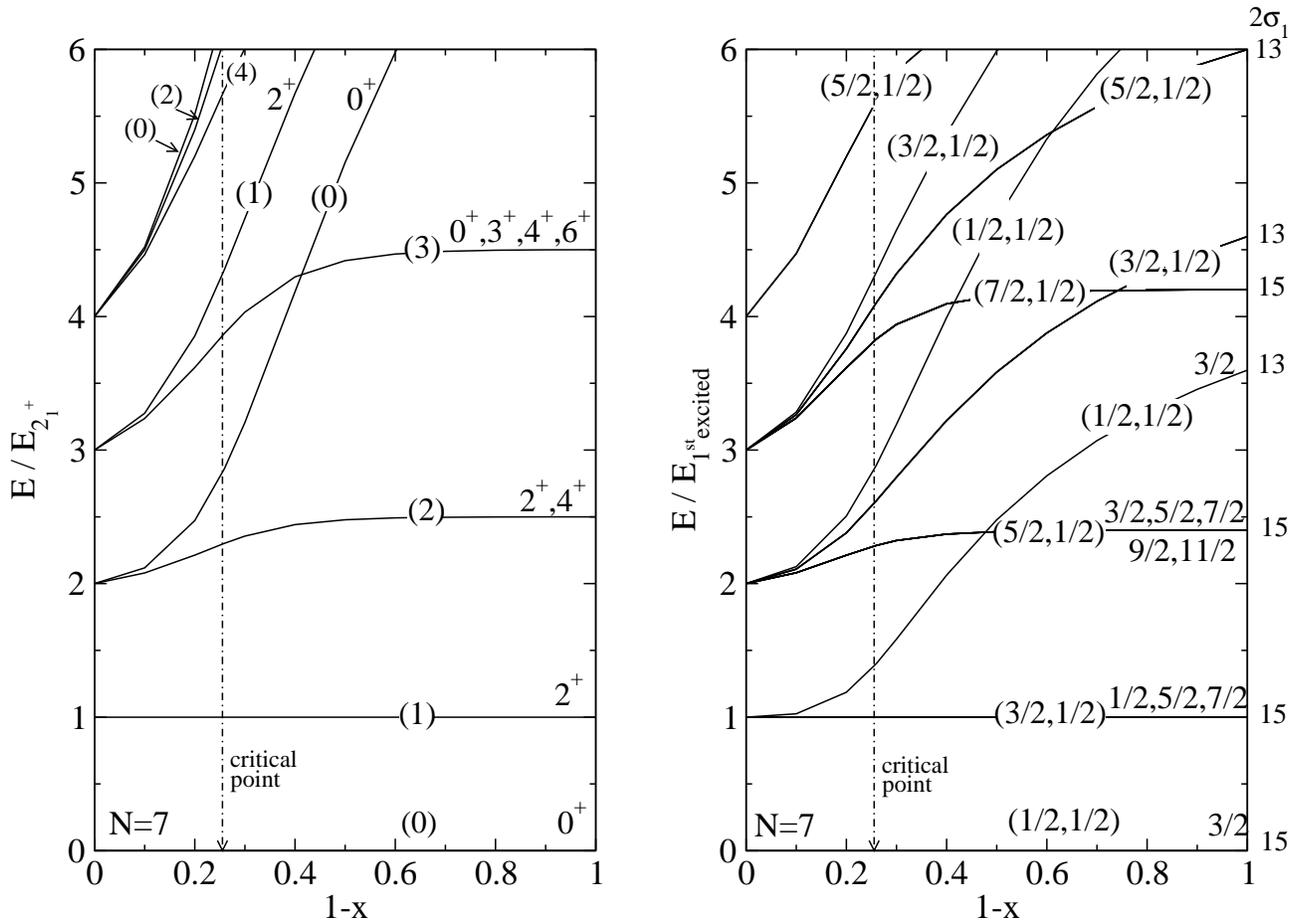}
\end{center}
\caption{ 
Energy levels (normalized to the energy of the first excited state) for the even 
and odd systems are displayed as a function of the parameter $(1-x)$ 
for the boson (\ref{HPP}) 
and boson-fermion (\ref{HBF}) Hamiltonians.  A number $N=7$ of bosons has been 
assumed in both cases, while the odd particle has been taken in the j=3/2 orbital.
In the left panel (even case) we indicate for each level the 
$\tau$ quantum number (in parenthesis), spin and parity.  In the right panel (odd case)
we quote the $(\tau_1,\tau_2)$ quantum numbers (in parenthesis) and spin.  In the
extreme $x=0$ case we also indicate the $\sigma_1$ quantum number. The position of 
the even critical point is marked.  
}
\label{fig1}
\end{figure}  

\begin{sidewaysfigure}[!t]
\begin{center}
\begin{picture}(500,340)(0,0)
\psset{xunit=2.6pt,yunit=190pt} 

\psline{-}(0,0)(190,0) \rput(200,0){  3/2}
\rput(-13,0){(1/2,1/2)}

\psline{-}(0,0.51)(190,0.51) \rput(200,0.51){  1/2}\rput(-13,0.55){(3/2,1/2)}
\psline{-}(0,0.55)(190,0.55) \rput(200,0.55){  5/2}
\psline{-}(0,0.59)(190,0.59) \rput(200,0.59){  7/2}

\psline{-}(0,1.18)(190,1.18) \rput(200,1.18){  3/2}\rput(-13,1.26){(5/2,1/2)}
\psline{-}(0,1.22)(190,1.22) \rput(200,1.22){  5/2}
\psline{-}(0,1.26)(190,1.26) \rput(200,1.26){  7/2}
\psline{-}(0,1.30)(190,1.30) \rput(200,1.30){  9/2}
\psline{-}(0,1.34)(190,1.34) \rput(200,1.34){  11/2}

\psline{-}(80,0.77)(190,0.77) \rput(200,0.77){  3/2}
\rput(-13,0.77){(1/2,1/2)}

\psline{-}(80,1.41)(190,1.41) \rput(200,1.41){  1/2}
\psline{-}(80,1.45)(190,1.45) \rput(200,1.45){  5/2}\rput(-13,1.45){(3/2,1/2)}
\psline{-}(80,1.49)(190,1.49) \rput(200,1.49){  7/2}

\psline{-}(160,1.60)(190,1.60) \rput(200,1.60){  3/2}
\rput(-13,1.60){(1/2,1/2)}  \rput(-13,1.70){$(\tau_1,\tau_2)$}

\rput(-26,0){0}\rput(-26,0.55){1}\rput(-26,0.77){1.40}\rput(-26,1.26){2.29}
\rput(-26,1.45){2.64}\rput(-26,1.60){2.91} \rput(-26,1.70){$\epsilon$}

\rput(210,0){1}\rput(210,0.55){1}\rput(210,0.77){2}\rput(210,1.26){1}\rput(210,1.45){2}
\rput(210,1.60){3}\rput(210,1.70){$\xi$}

\psline{->}(2,0.59)(2,0)
\psline{->}(5,0.55)(5,0) 
\psline{->}(8,0.51)(8,0) 
\psline[linewidth=18pt,linecolor=white](0,.25)(10,.25)\rput(4,0.25){  100}
\psline{->}(11,0.59)(11,0.55)  \rput{0}(6,0.57){  19.2}
\psline{->}(14,0.55)(14,0.51)  \rput{0}(10,0.53){  13.1}

\psline{->}(5,1.34)(5,1.30)  \rput{0}(0,1.32){  12.7}
\psline{->}(5,1.30)(5,1.26) \rput{0}(0,1.28){  0.13}
\psline{->}(5,1.26)(5,1.22) \rput{0}(0,1.24){  8.49}
\psline{->}(5,1.22)(5,1.18) \rput{0}(0,1.20){  2.29}

\psline{->}(11,1.34)(11,1.26)  \rput{90}(11,1.39){  0.61}
\psline{->}(14,1.30)(14,1.22) \rput{90}(14,1.38){  5.7}
\psline{->}(17,1.26)(17,1.18) \rput{90}(17,1.38){  4.4}

\psline{->}(20,1.34)(20,0.59) 
\psline{->}(23,1.30)(23,0.59) 
\psline{->}(26,1.30)(26,0.55)
\psline{->}(29,1.26)(29,0.59)
\psline{->}(32,1.26)(32,0.55) 
\psline{->}(35,1.22)(35,0.59)
\psline{->}(38,1.22)(38,0.55) 
\psline{->}(41,1.22)(41,0.51) 
\psline{->}(44,1.18)(44,0.59)
\psline{->}(47,1.18)(47,0.55) 
\psline{->}(50,1.18)(50,0.51)
\psline[linecolor=white,linewidth=28pt](18,.80)(55,.80)
\rput{90}(20,0.80){  141.1} \rput{90}(23,0.80){  30.2}
\rput{90}(26,0.80){  110.9} \rput{90}(29,0.80){  69.1}
\rput{90}(32,0.80){  72.0}  \rput{90}(35,0.80){  34.8}
\rput{90}(38,0.80){  21.6}  \rput{90}(41,0.80){  84.6}
\rput{90}(44,0.80){  16.1}  \rput{90}(47,0.80){  75.6}
\rput{90}(50,0.80){  49.4}

\psline{->}(82,1.49)(82,0.77) 
\psline{->}(85,1.45)(85,0.77) 
\psline{->}(88,1.41)(88,0.77) 
\psline[linewidth=18pt,linecolor=white](80,1.)(90,1.)\rput(85,1.){ 81.6}
\psline{->}(91,1.49)(91,1.45)  \rput{0}(86,1.47){  13.1}
\psline{->}(94,1.45)(94,1.41)  \rput{0}(90,1.43){  8.9}

\psline{->}(92,0.77)(72,0.59) \rput{90}(92,0.82){  1.02}
\psline{->}(95,0.77)(75,0.55) \rput{90}(95,0.82){  0.76}
\psline{->}(98,0.77)(78,0.51) \rput{90}(98,0.82){  0.25}

\psline{->}(102,0.77)(82,0.) \rput{70}(87,0.28){  22.7}

\psline{->}(101,1.49)(94,1.34) \rput{90}(99,1.55){ 0.95}
\psline{->}(103,1.49)(96,1.30) \rput{90}(102,1.55){ 0.20}
\psline{->}(105,1.49)(99,1.26) \rput{90}(105,1.55){ 0.37}
\psline{->}(107,1.49)(102,1.22)\rput{90}(108,1.55){0.14}
\psline{->}(109,1.49)(105,1.18)\rput{90}(111,1.55){0.04}
\rput(119,1.5){\psframebox*{~~~~~~~~~}}
\psline{->}(117,1.45)(110,1.30)  \rput{90}(115,1.5){1.0}
\psline{->}(119,1.45)(113,1.26) \rput{90}(118,1.5){ 0.49}
\psline{->}(121,1.45)(116,1.22)\rput{90}(121,1.5){  0.12}
\psline{->}(123,1.45)(119,1.18) \rput{90}(124,1.5){0.27}
\psline{->}(127,1.41)(124,1.22) \rput{90}(128,1.55){ 1.37}
\psline{->}(129,1.41)(127,1.18) \rput{90}(131,1.55){ 0.54}

\psline{->}(135,1.49)(120,0.59) 
\psline{->}(137,1.49)(122,0.55) 
\psline{->}(139,1.45)(126,0.59)
\psline{->}(141,1.45)(128,0.55) 
\psline{->}(143,1.45)(130,0.51) 
\psline{->}(146,1.41)(134,0.55) 
\psline[linewidth=18pt,linecolor=white](120,.82)(140,.82)
\rput{90}(123,.82){5.6} 
\rput{90}(126,.82){5.4} 
\rput{90}(129,.82){6.7}
\rput{90}(132,.82){0.15}
\rput{90}(135,.82){3.64}
\rput{90}(138,.82){10.9}
\psline{->}(154,1.49)(132,0.) 
\psline{->}(156,1.45)(136,0.) 
\psline{->}(158,1.41)(140,0.) 
\psline[linewidth=18pt,linecolor=white](143,1.)(155,1.)\rput(149,1.){ 0.24}

\psline{->}(160,1.60)(160,1.49) \rput{90}(158,1.56){  0.83}
\psline{->}(165,1.60)(165,1.45) \rput{90}(163,1.545){  0.62}
\psline{->}(170,1.60)(170,1.41) \rput{90}(168,1.53){  0.21}

\psline{->}(176,1.60)(156,0.77) \rput{70}(158,0.92){  13.7}

\psline{->}(181,1.60)(159,0.59)
\psline{->}(183,1.60)(163,0.55) 
\psline{->}(185,1.60)(167,0.51)
\psline[linewidth=18pt,linecolor=white](164,1.)(184,1.)
\rput{90}(168,1.0){14.8} \rput{90}(171.5,1.0){11.1} \rput{90}(175,1.0){3.7}

\psline{->}(189,1.60)(179,0.) \rput{83}(182,0.83){  0.51}

\rput(200,1.70){j}

\end{picture}
\caption{Energy levels and quadrupole transition rates $B(E2\downarrow)$
for the odd system at the core critical point.  For illustration
purposes the various multiplets, labeled by the $(\tau_1,\tau_2)$ quantum numbers in 
parenthesis, have been arbitrarily split according to their $j$ quantum number. 
The label on the extreme left is the energy in relative units, while the 
label at
the extreme right corresponds to the
label $\xi$, used in \cite{I05}.
$B(E2\downarrow)$'s have been normalized to the value 100 for the  
transitions (with equal strengths)
between the states of the first multiplet and the ground state.
}
\end{center}
\label{fig2}
\end{sidewaysfigure}
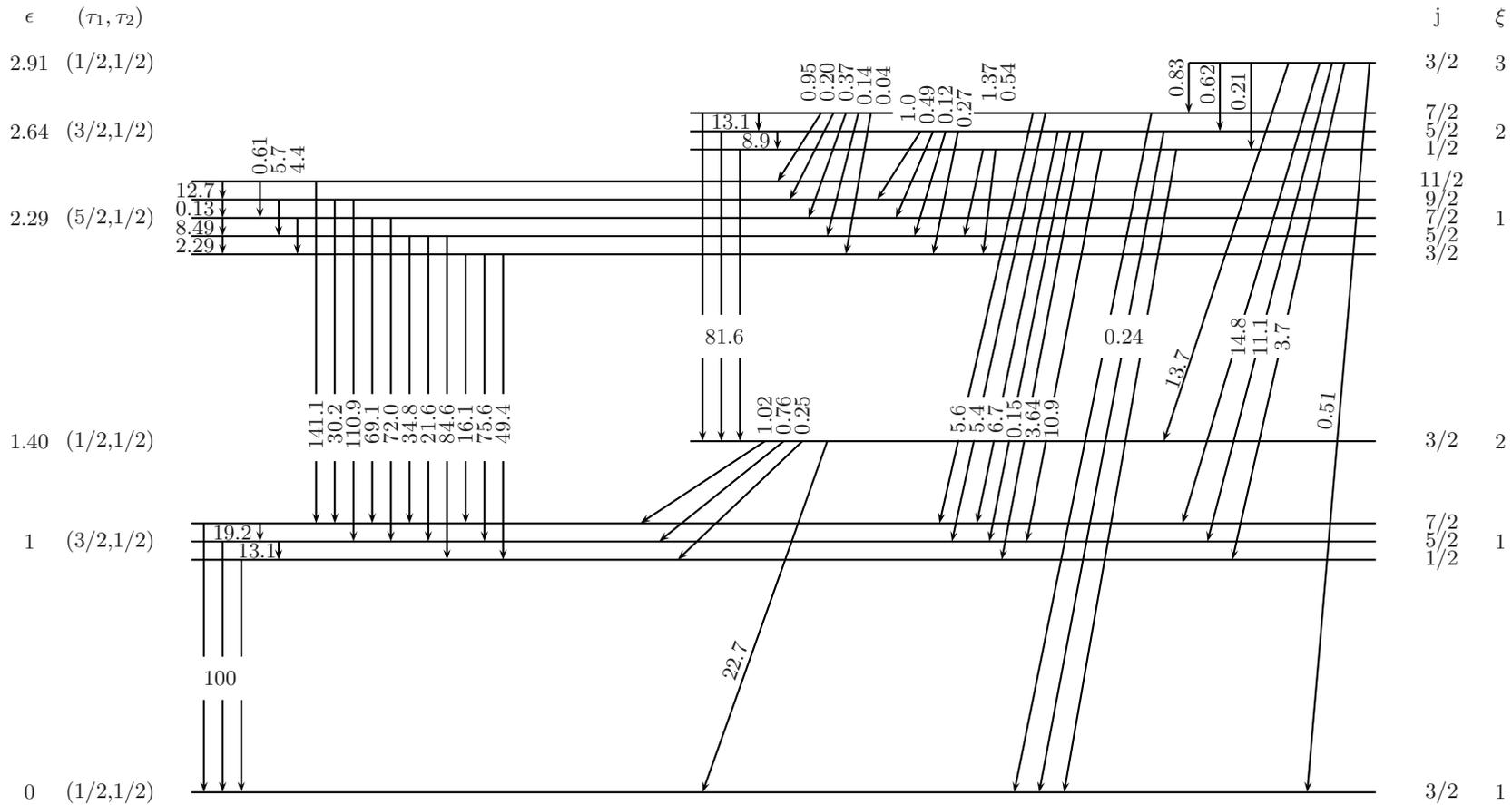

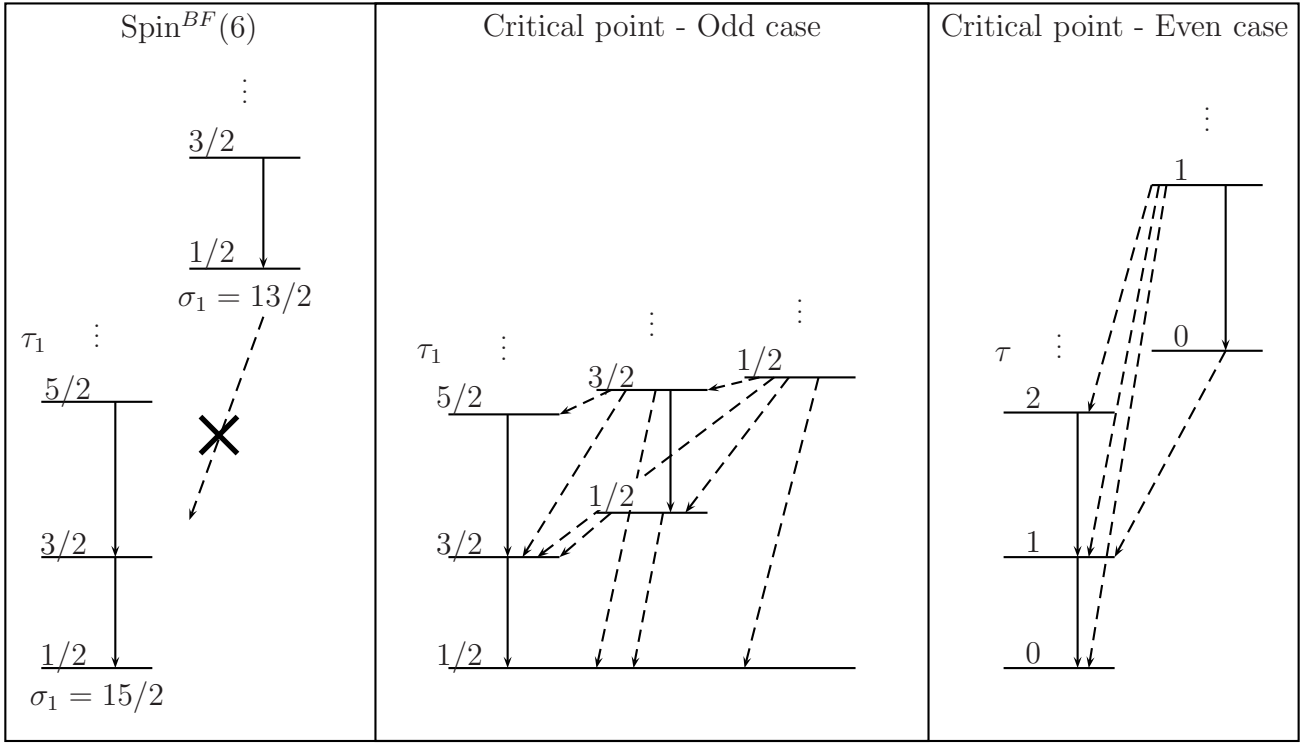
\begin{figure}[!t]
\begin{center}
\vspace{3.6cm}
\begin{picture}(350,200)(80,0)
\psset{unit=1.4pt}
\psframe(0,0)(350,200)
\psframe(100,0)(250,200)

\psline{-}(10,20)(40,20) \rput(16,23){\large $1/2$} 
\rput(25,12){\large  $\sigma_1=15/2$}
\psline{-}(10,50)(40,50)\rput(16,53){\large  $3/2$}
\psline{-}(10,92)(40,92) \rput(16,95){ \large $5/2$} \rput(8,108){\large $\tau_1$}
\rput(25,112){$\vdots$}
\rput(65,120){\large  $\sigma_1=13/2$}
\psline{-}(50,128)(80,128)\rput(56,132){\large  $1/2$}
\psline{-}(50,158)(80,158)\rput(56,162){\large  $3/2$}
\rput(65,178){$\vdots$}
\rput(50,193){\large Spin$^{BF}$(6)}

\psline{-}(120,20)(230,20)\rput(123,23){\large  $1/2$}
\psline{-}(120,50)(150,50)\rput(123,53){\large  $3/2$}
\psline{-}(120,88.6)(150,88.6)\rput(123,92.6){\large  $5/2$} \rput(115,105){\large $\tau_1$}
\rput(135,108.6){$\vdots$}
\psline{-}(160,62)(190,62)
\psline{-}(160,95.2)(190,95.2) \rput(164,98.2){\large  $3/2$}
\rput(175,115.2){$\vdots$}
\psline{-}(200,98.6)(230,98.6)\rput(204,102.6){\large  $1/2$}
\rput(215,118.6){$\vdots$}
\rput(175,193){\large Critical point - Odd case}

\psline{-}(270,20)(300,20)\rput(278,24){\large  $0$}
\psline{-}(270,50)(300,50) \rput(278,54){\large  $1$}
\psline{-}(270,89.1)(300,89.1) \rput(278,93.1){\large  $2$}\rput(270,105){\large $\tau$}
\rput(285,109.1){$\vdots$}
\psline{-}(310,105.8)(340,105.8)\rput(318,109.8){\large  $0$}
\psline{-}(310,150.6)(340,150.6)\rput(318,154.6){\large  $1$}
\rput(325,170.6){$\vdots$}
\rput(300,193){\large Critical point - Even case}

\psline{->}(30,50)(30,20)\psline{->}(30,92)(30,50)\psline{->}(70,158)(70,128)
\psline[linestyle=dashed]{->}(70,115)(50,60)
\psline[linewidth=2pt]{-}(53,88)(63,78)\psline[linewidth=2pt]{-}(53,78)(63,88)

\psline{->}(136,50)(136,20)\psline{->}(136,88.6)(136,50)
\psline{->}(180,95.2)(180,62)\psline[linestyle=dashed]{->}(164,62)(150,50)
\psline[linestyle=dashed]{->}(178,62)(170,20)
\psline[linestyle=dashed]{->}(176,95.2)(160,20)
\psline[linestyle=dashed]{->}(164,95.2)(150,88.6)
\psline[linestyle=dashed]{->}(168,95.2)(140,50)
\psline[linestyle=dashed]{->}(220,98.6)(200,20)
\psline[linestyle=dashed]{->}(204,98.6)(190,95.2)
\psline[linestyle=dashed]{->}(208,98.6)(144,50)
\psline[linestyle=dashed]{->}(212,98.6)(184,62)

\psline{->}(290,50)(290,20)\psline{->}(290,89.1)(290,50)
\psline{->}(330,150.6)(330,105.8)
\psline[linestyle=dashed]{->}(330,105.8)(300,50)
\psline[linestyle=dashed]{->}(314,150.6)(293,20)
\psline[linestyle=dashed]{->}(312,150.6)(293,50)
\psline[linestyle=dashed]{->}(310,150.6)(293,89.1)

\psline[linewidth=12pt,linecolor=white]{-}(158,67)(178,67)
\rput(164,66){\large  $1/2$}

\end{picture}
\caption{Schematic illustration of allowed quadrupole transitions for the odd
system  at the $Spin^{BF}(6)$ (left panel) and critical point (middle panel) cases, 
and for the even case at the critical point (right panel). Solid lines indicate
stronger in-band transitions with respect to the inter-band transitions shown as 
dashed lines. Horizontal lines correspond to multiplets including several j values. 
}
\end{center}
\label{fig3}
\end{figure}

\begin{figure}[!t]
\begin{center}
\epsfig{file=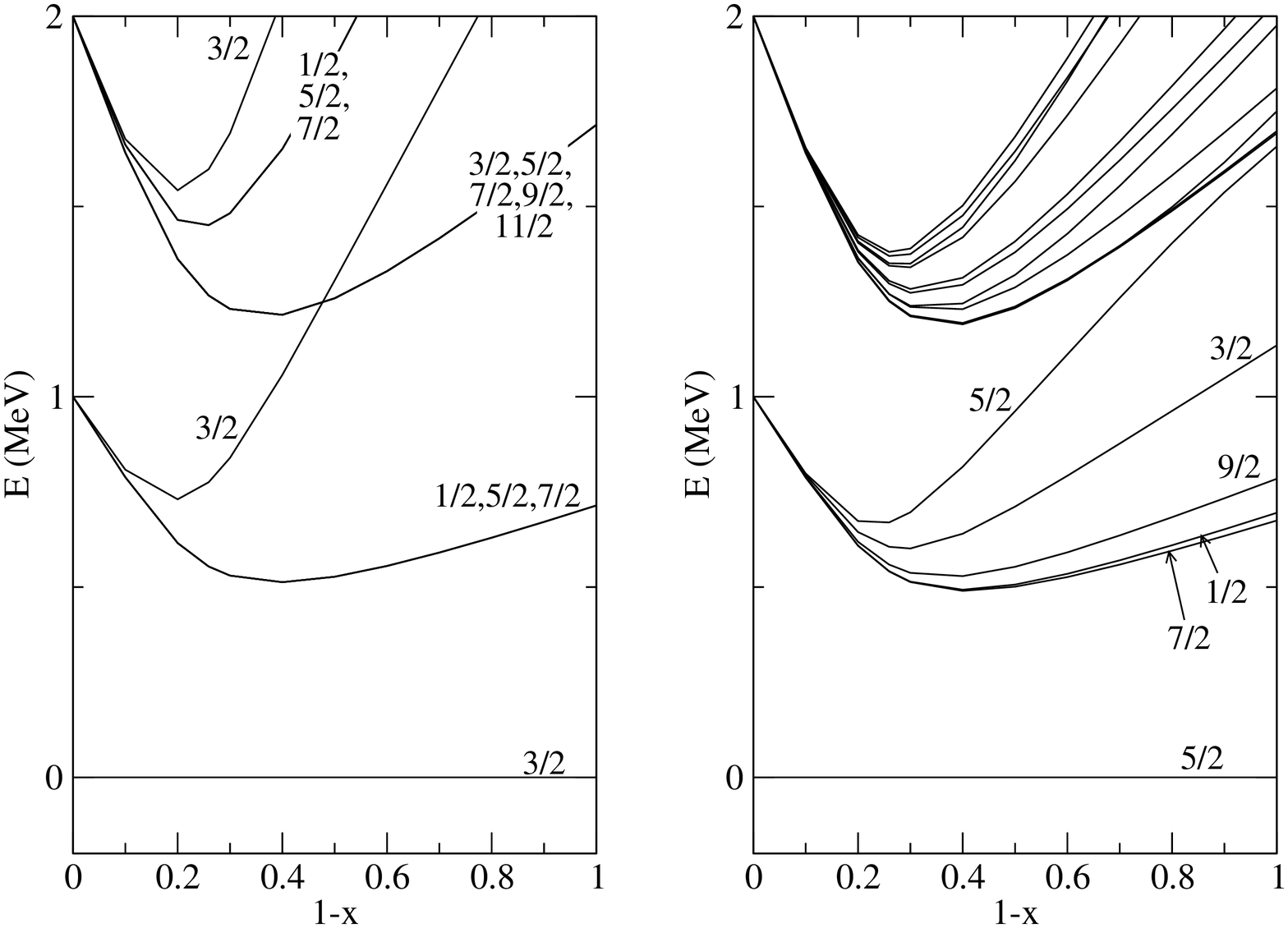,clip=,width=0.95\textwidth}
\end{center}
\caption{ 
Energy levels for the odd system are displayed as a function of the 
parameter $(1-x)$ 
for the boson-fermion Hamiltonian (\ref{HBF}).  A number $N=7$ of bosons has been 
assumed, while the odd particle has been taken in the j=3/2 orbital (left panel) 
and j=5/2 orbital (right panel).
The spin quantum number is indicated for each state.}
\label{fig4}
\end{figure}

This choice of the fermion space and the boson-fermion interaction is such that
one recovers, in the extreme cases, the Bose-Fermi symmetry\cite{IVI91} associated with 
the $Spin^{BF}(6)$ group (for $x$=0) and the vibrational $U^B(5) \otimes SU^F(4)$ case 
(for $x$=1). In analogy with the overall O(5) structure in the even
case, this selection guarantees the preservation of the degeneracies associated 
with the $Spin^{BF}(5)$ symmetry for any value of $x$. This can be
more clearly seen within the intrinsic frame formalism \cite{GK80,DSI80,BM80}.  
In order to get energy eigenvalues for the case of a $j=3/2$
particle coupled to the ground state  $|\Phi_B(\beta,\gamma) \rangle$
boson condensate one has to diagonalize the Hamiltonian in a space of
dimension 4 with basis vectors: $|\Phi_B(\beta,\gamma) \rangle \otimes |j,m\rangle$. 
The pure boson part for the case considered here is only a function of $\beta$
and gives a global diagonal contribution. The
pure fermionic part is just a constant. Thus, the only remaining part,
$V_{BF}$, has to be diagonalized. Its eigenvalues, doubly degenerate,
turn out to be $\gamma$-independent,
\begin{equation}
E_{\pm}(\beta,\gamma)~=E_{\pm}(\beta)~=~\pm 2\frac{(1-x)\beta}{1+\beta^2} ~~. 
\end{equation}
In other words, the addition of the odd particle does not destroy the 
$\gamma$-instability of the system, giving rise to energy surfaces for the different 
odd intrinsic states that are still $\gamma$-independent \cite{AA92}.  The 
particular behaviour of the j=3/2 orbital was first put in evidence
by Bayman and Silverberg \cite{BS60} within the collective model. 

The resulting energy spectra in the odd system are shown in the right
panel of Fig.\ref{fig1}
as a function of the control parameter
$1-x$.  The total number of bosons, $N$, has been assumed to
be equal to 7.
For a better comparison, we also show in the left panel of the figure the 
corresponding evolution of the spectrum in the even core. It should be remembered
that the use of a finite number of bosons does not lead to the same results 
as the corresponding ones obtained in the equivalent situation within the Bohr 
Hamiltonian, which is only reached in the limit of infinite number 
of bosons.  Note also that the geometrical limit obtained from the boson Hamiltonian
(\ref{HPP}) for large values of $N$ corresponds to the case of
the collective potential behaving as $\beta^4$, and not precisely to the E(5) case, 
that corresponds to the infinite square well.
The level evolution in 
the odd case shows a behaviour qualitatively similar to that of the even case.
The group structure of $Spin^{BF}(5)$ with respect to O(5) simply leads to a richer
pattern for the fermion case and slightly different ratios for the energy levels.
For example in the limiting $x$=0 case, corresponding to $Spin^{BF}(6)$ and O(6)
symmetry groups for the odd and even nuclei, respectively, we obtain for the ``ground bands'',
with
maximum $\sigma$
values ($\sigma_1=N+1/2$ and $\sigma=N$ for odd and even nuclei, respectively), the ratios
$E(\tau_1=5/2,\tau_2=1/2)/E(\tau_1=3/2,\tau_2=1/2)=2.4$ and
$E(\tau_1=7/2,\tau_2=1/2)/E(\tau_1=3/2,\tau_2=1/2)=4.2$ with respect
to the values
$E(\tau=2)/E(\tau=1)=2.5$ and $E(\tau=3)/E(\tau=1)=4.5$ of the even case.

As in the even case, the levels that in the extreme $x=0$ limit eventually 
correspond to the 
``ground band'' ($\sigma_1=N+1/2$) show a rather smooth (or even flat) behaviour, aside from the
changes around the critical point. Instead, the levels that will end up 
with other values of $\sigma_1$
(as the levels corresponding
to smaller values of $\sigma$  
in the even case) show a more violent variation in the whole range. A typical case
is the second $(\tau_1=1/2,\tau_2=1/2)j=3/2$ state that starts at one-phonon energy
in the vibrational limit to end up as the third j=3/2 state in the opposite limit.
The position of this 3/2 state is the key element to characterize the particular
situation and its position along the transitional path.  The position of this state 
plays the same role as the key position of the first excited $0^+$ state in even nuclei.
The smoother behaviour of the ground band levels with respect to the other bands 
confirms the fact that to establish a definite critical situation it is not 
at all sufficient to rely just on the sequence of energy levels in the ``ground'' 
band, and that some of the claims for the occurrence of definite transitional 
symmetries have to be taken with serious caution.



Crucial and selective information on nuclear spectra comes from the 
transition probabilities, in particular from the electric quadrupole ones.  
These are obtained in terms of the matrix elements of the electric
quadrupole operator.  In our IBFM case this operator
is given by
\begin{equation}
\hat Q_{BF}=\hat Q_B+\hat q_F= 
(s^\dag \times \tilde d + d^\dag \times \tilde s)^{(2)} +
 (a^\dag _{j} \times \tilde a_{j})^{(2)}.
\label{QBF}
\end{equation} 
Note that, consistently with the boson quadrupole operator used in the Hamiltonian,
we have not included any $(d^\dag \times \tilde d)^{(2)}$ term in the transition 
operator.

Values of the individual transition probabilities, state by state, for the
odd nucleus at the critical point situation are shown in Figure 2.
In it, we have plotted the lowest 6 multiplets, which have been arbitrarily
split in order to show the possible E2 transitions between the different states. In
the figure the levels have different lengths depending on the $\xi$ value,
which labels diferent families and it is shown along with the spin of the state 
(we conform here to the notation introduced in \cite{I05}). Although the general 
pattern remains the same, the detailed energy sequence depends on the number of
interacting bosons, $N$. Owing to the properties of
the $Spin^{BF}(5)$ group, the $\Delta\tau_1=0,\pm 1, \Delta\tau_2=0$
selection rule still hold. 
It can be observed that E2 transitions are stronger between 
states with the same $\xi$ value and $\Delta\tau_1= 1$. Transitions 
between states pertaining to families with different $\xi$ are 
one or two orders of magnitude smaller. Transitions 
between states with the same $\xi$  and $\tau_1$ values, but 
different spins, corresponding to the same multiplets are also one or two orders of magnitude 
smaller than the ones between different multiplets in the same band.

 The allowed quadrupole transitions are sketched in Figure 3 for the odd system for
the $Spin^{BF}(6)$ case
and  for the critical point situation, and for the even case at the critical point. 
In addition
to the difference in the level sequence, in the second case transitions are allowed
between different bands (although weaker than inband transitions, as already mentioned),
in a similar way as inter-band transitions are allowed
at the critical point for the even case
(as for example between the second $0^+$, with $\tau=0$, and the first $2^+$, with  $\tau=1$, 
as shown in the figure). 

The comparison of our critical point solution with Iachello's one evidences a similar overall 
organization of the spectrum, although with sizeable differences in the relative position of 
the different $\xi$ bands. It must be recalled that, in
principle, we should compare the E(5/4) solution with the large $N$ limit of the interacting 
boson model, while we have concentrated, along this paper, on the $N=7$ case. 
The first $\tau_1=5/2$ multiplet in the E(5/4) lies at around 2.20, close to the value 2.29 
obtained in our model. On the other hand, the two first multiplets 
of the $\xi=2$ family of excited states of the E(5/4) solution lie at around 3.33 and 5.02, 
compared with the corresponding values 1.40 and 2.64 of our
model. This inversion of the position of the $\xi=1,\tau_1=5/2$ and $\xi=2,\tau_2=1/2$
multiplets is the first evident difference between the two models. Another important
difference is seen in the relative B(E2) values: although in-band transitions display
comparable values, significant discrepancies are observed in interband transition 
(notably the transitions between the $\xi=2,\tau_1=1/2$ state and the two lowest multiplets
of the ground state band).

Preserving the degeneracies associated with the $Spin^{BF}(5)$ symmetry is not
an exclusive property of the j=3/2 orbitals.  Other cases are known in the
literature, for example the case of the odd particle moving in the j=1/2,3/2,5/2
orbitals, under the condition of special values for the fermion quadrupole 
matrix elements.  Also in this case one recovers, for the $x$=0 limiting case,
the Bose-Fermi symmetry behaviour.  But this does not happen in more general cases
with other combinations of orbitals or fermion quadrupole matrix elements 
\cite{Jo2004}.  To give an idea,
we present in Figure 4 the evolution of the spectrum as a function of the 
control parameter $1-x$ in the cases of a single j=3/2 and j=5/2. 
The figure shows a spectrum for j=5/2 (right panel) that is qualitatively 
similar to
the one displayed in the left panel for j=3/2, but is more complex due to the 
removal of all degeneracies.

To summarize, we have considered, within the Interacting Boson Fermion Model, the 
coupling of an odd j=3/2 particle to a boson
core that undergoes a transition from spherical U(5) to
$\gamma$-unstable O(6) character. The particular choice of the Hamiltonian and
of the  j=3/2 orbital
preserves in the odd case the condition of $\gamma-$instability of the system, and it is 
reflected in the preservation of the degeneracies associated with the
$Spin^{BF}(5)$ symmetry. As a consequence, the energy spectrum and the electromagnetic 
transitions for the odd nucleus with a critical core
display behaviours 
qualitatively similar to those characterizing the phase transition in the
even core. We have compared our results with the recently proposed E(5/4) 
approach, based on the Bohr hamiltonian. Both approaches display similar
qualitative pictures, although we evidence a number of quantitative differences,
that can be traced back to the different nature of the two schemes.

\section*{Acknowledgements}

This work was supported by the Italian-Spanish agreement INFN-CICYT and 
by the Spanish DGICYT under project number FIS2005-01105. 
We acknowledge enlightening discussions with J. Barea and B.F. Bayman.

\end{document}